\pdfoutput=1
\documentclass[journal]{IEEEtran}

\usepackage{amsmath,graphicx}
\graphicspath{{./img/}}

\usepackage{enumitem}

\usepackage{algorithm} 
\usepackage{algorithmic}

\usepackage{amssymb}
\usepackage{amsfonts}
\usepackage{amsthm}
\usepackage{booktabs}
\usepackage{url}
\usepackage{mathtools}
\usepackage{cite}

\usepackage[caption=false,font=footnotesize]{subfig}

\usepackage{color}
\usepackage{xcolor}

\usepackage{soul}
\usepackage{todonotes}
\usepackage[utf8]{inputenc}
\usepackage[T1]{fontenc}

\newcommand{\minimize}{\operatornamewithlimits{minimize}}

\newcommand{\subjto}{\operatornamewithlimits{\text{subject to}}}

\title{Range and Bearing Data Fusion for Precise Convex Network Localization}

\author{Cl\'audia Soares*, Filipa Valdeira, and Jo\~ao
  Gomes,~\IEEEmembership{Member,~IEEE} \thanks{This work has received
    funding from the European Union’s Horizon 2020 research and
    innovation programme under the Marie Skłodowska-Curie grant
    agreement No 812912, and from FCT Portugal (UID/EEA/50009/2019 and
    HARMONY PTDC/EEI-AUT/31411/2017)
    C. Soares and J. Gomes are with the Institute for Systems and
    Robotics, Instituto Superior T\'ecnico, Universidade de Lisboa (\{csoares,jpg\}@isr.tecnico.ulisboa.pt) and F. Valdeira
    is with the University of Milan (filipa.marreiros@unimi.it).}  }

\begin{document}
%




\maketitle

% Countless systems require localization methods to accomplish their
%   proposed tasks. While GNSS solutions offer great performance, for
%   large networks this may not be affordable, as it depends on
%   expensive instrumentation. Besides, environmental constraints can
%   even prevent such approaches, as it is the case in indoor or
%   underwater settings. Most solutions so far avoid the
%   maximum-likelihood estimator

\begin{abstract}
  Hybrid localization in GNSS-challenged environments using measured
  ranges and angles is becoming increasingly popular, in particular
  with the advent of multimodal communication systems. Here, we
  address the \emph{hybrid network localization
    problem} using ranges and bearings to jointly determine the
  positions of a number of agents through a \emph{single
    maximum-likelihood} (ML) optimization problem that seamlessly
  fuses all the available pairwise range and angle measurements. We
  propose a tight convex surrogate to the ML estimator, we examine
  practical measures for the accuracy of the relaxation, and we
  comprehensively characterize its behavior in simulation. We found
  that our relaxation outperforms a state of the art SDP relaxation by
  one order of magnitude in terms of localization error, and is
  amenable to much more lightweight solution algorithms.
\end{abstract}

%\begin{IEEEkeywords}

%\end{IEEEkeywords}

% \fi
\IEEEpeerreviewmaketitle

\section{Introduction}

%MOTIVATION
Spatial awareness is a hallmark of contemporary real-world systems and
applications, particularly when multiple agents collaborate to attain
common goals. Location technologies are key for operation in
GNSS-challenged environments, like
underwater~\cite{PaullSaeediSetoLi2014} or indoor~\cite{Schneider2013}
and city skyscraper areas~\cite{Moustaka2019}, and also in many
applications in wireless communications \cite{Witrisal2016}, sensor
networks \cite{Buehrer2018}, IoT~\cite{Turjman2017},
medicine~\cite{Hossain2017,DeyEtAl2017}, etc. Our work addresses the
network localization problem~\cite{AspnesErenGoldenbergAnderson2006},
where multiple networked agents cooperate in sensing and computation
to jointly estimate their unknown positions.
% This is more challenging than GPS-like or radar-like single-source
% localization~\cite{BeckStoicaLi2008}.

\paragraph*{Related work}
\label{sec:related}
% In network localization agents can rely on distance and/or angle
% measurements between themselves and a few landmarks to estimate their
% positions.
There is a vast literature for range-based network localization, from
multidimensional scaling\cite{ji2004,jiang2016} to nonconvex ML
estimation~\cite{SoaresXavierGomes2014a,Zhou2018}, convexifications of
ML problems~\cite{SimonettoLeus2014,soares2014simple,Piovesan2016},
and other convex heuristics to match measured data to a data
model~\cite{ShiHeChenJiang2010,Saab2016}.

Several current technologies not only give access to accurate distance
measurements, but also provide angle information. These added
measurements can improve the quality of estimates or reduce the amount
of resources spent to obtain a reasonable localization precision. 5G
is an especially noteworthy example where the value of hybrid
measurements has been noted~\cite{Agiwal2016, Witrisal2016}, and
information-theoretic bounds are available to characterize their
impact~\cite{Shen2010, Buehrer2018}. While the topic of single-source
range/bearing localization is reasonably well covered in the technical
literature, specific references addressing \emph{network}
localization algorithms for that data model are surprisingly
scarce~\cite{Patwari2005,Biswas2005,eren2011,FerreiraGomesCosteira2015,FerreiraGomesSoaresCosteira2016,Naseri2018}
(see also the recent survey~\cite{Buehrer2018}). Yet, the potential
usefulness of such algorithms, e.g., in massive MIMO communication
scenarios seems quite obvious. Below, we focus our literature review
on the subclass of hybrid network localization algorithms derived from
single convex formulations, which avoid initialization issues
affecting other types of approaches.
%% Go through all {Patwari2005,eren2011,FerreiraGomesCosteira2015,FerreiraGomesSoaresCosteira2016,Naseri2018}
The importance of considering hybrid measurements was emphasized early
on in~\cite{Patwari2005}, but a clear statement of the graphical
conditions for localizability of the network assuming range and
bearing measurements was only formalized in~\cite{eren2011}. One of
the first convex formulations was a semidefinite program (SDP)
proposed in \cite{Biswas2005} for 2D scenarios. Angle constraints were
manipulated into a form similar to the one used for ranges, and
incorporated into an existing range-only
SDP. Reference~\cite{Ferreira2018} addresses the single-source and
network localization problems through a convex relaxation of a
nonconvex least-squares cost function,
and~\cite{FerreiraGomesSoaresCosteira2016} extends this to mobile
setups. The very recent work in~\cite{wu2019cooperative} explores the
problem using belief propagation, but relies on linearized
approximations. Recently,~\cite{Naseri2018} takes the ML estimator for
the original static scenario and considers Gaussian noise for ranges
and von Mises--Fisher noise for bearings. This very interesting work
performs several approximations and formulates the problem as an
SDP. However, the manipulations involve squaring of range and angular
terms, which is known to amplify noise and degrade localization
accuracy~\cite{kay1993,OguzGomesXavierOliveira2011}.

Another important line of work for hybrid network localization,
particularly in the scope of wireless communications, uses RSS-based
measurements as proxies for ranges (see~\cite{Tomic2019,Tomic2017} for
an extensive list of references). The model for RSS measurements is
quite different from the one that we adopt for ranges, and so are the
manipulations and relaxations used in localization algorithms.

\paragraph*{Contributions}
\label{sec:contribs}

As in~\cite{Naseri2018} we adopt a ML approach assuming Gaussian noise
for range measurements and von Mises--Fisher noise for bearings,
leading to a difficult to solve nonconvex problem.
Unlike~\cite{Naseri2018} we do not approximate the problem via
squaring of range or angle terms, but instead adopt an unconventional
relaxation technique that in our simulations attains one order of
magnitude more accurate results. The approach works in 2D and 3D (or
in any ambient dimension). As a second contribution, we provide
certificates of optimality that indicate if the minimizer of the
convex surrogate coincides with that of the nonconvex ML
estimator. While our formulation is amenable to parallelization, the
derivation of tailored solution algorithms is beyond the scope of this
paper. Our main goal here is to highlight and characterize the
excellent accuracy of the proposed approximate ML relaxation.

\section{Problem statement}
\label{sec:problem}
We model the network of agents as a
graph~${\mathcal G}=({\cal V},{\cal E})$ where
${\cal V} = \{1,\cdots, n\}$ denotes the set of~$n$ agents with
unknown positions, and each edge~$i \sim j \in {\cal E}$ indicates
that agents~$i$ and~$j$ can communicate, and can measure a noisy
version of their scalar distance~$d_{ij}$, that we consider
symmetrical. We express the position of agent~$i$ in the ambient space
as~$x_i \in \mathbb{R}^p$, where~$p \in \{2,3\}$ in practical
applications. We also define an edge set~${\cal E}_u \in {\cal E}$ as
the set of edges providing noisy angle measurements between
agents. Bearing data is measured as a unit vector~$u_{ij}$ expressed
in the world frame. For localization in a global reference frame, the
problem assumes the existence of a set~${\cal A}$ of landmarks or
anchors whose absolute positions~$\{a_k\}_{k \in {\cal A}}$ are known.
% In our formulation we assume that such elements have no computing power.
Each agent~$i$ can measure noisy ranges~$\{r_{ik}\}$ for
all~$k \in {\cal A}_i\subseteq {\cal A}$ and, possibly,
bearings~$\{v_{ik}\}_{k \in {\cal U}_i}$,
where~${\cal U}_i \subseteq {\cal A}_i$ is the subset of anchors
reachable from node~$i$ that also provide bearing measurements. We
note that these assumptions differ from those of~\cite{Tomic2019},
e.g., where \emph{every} range-like (RSS) measurement needs a matching
angular measurement. In some settings ranging devices are much cheaper
than those used to measure angles\cite{PaullSaeediSetoLi2014}, so
requiring the latter only in a subset of nodes may be desirable.

\paragraph*{Noise model}
\label{sec:noise-model}
We model noisy range measurements $d_{ij}$, $r_{ik}$ as independent
and identically distributed (iid) normal random variables centered at
the true ranges with standard deviations~$\sigma_{ij}$
and~$\varsigma_{ik}$, respectively. Similarly, we model noisy bearings
as iid von Mises-Fisher random variables, independent from ranges,
centered at the true bearings with concentration
parameters~$\kappa_{ij}$,~$\varkappa_{ik}$.  In the following section
we formulate the ML estimator for the positions of all the
agents~$x = \{x_i\}_{i=1}^n$ as a nonconvex optimization problem.

\paragraph*{Maximum likelihood localization with distance and angle
  measurements}
\label{sec:ml-loc}
Assuming the noise models discussed above, we can
write the maximum likelihood estimator for the positions~$x$ of the
overall network as
\begin{equation}
  \label{eq:noncvx-mle}
  \minimize_{x} \; f(x) + f_u(x),
\end{equation}
where~$f$ represents the range-related terms, and~$f_u$ represents
the bearing terms. Specifically, we have, as~$f(x)$,
% \begin{equation}
%   \label{eq:noncvx-ranges}
%   \begin{split}
%     f(x) = &\sum_{i \sim j \in {\cal E}}\frac{1}{\sigma_{ij}^2}(\|x_{i}-x_{j}\|-d_{ij})^2\\
%     & + \sum_{i}\sum_{k \in {\cal A}_i} \frac{1}{\varsigma_{ik}^2}(\|x_{i}-a_{k}\|-r_{ik})^2,
% \end{split}
% \end{equation}
\begin{equation}
  \label{eq:noncvx-ranges}
     \sum_{i \sim j \in {\cal E}}\frac{1}{\sigma_{ij}^2}(\|x_{i}-x_{j}\|-d_{ij})^2
     + \sum_{i}\sum_{k \in {\cal A}_i} \frac{1}{\varsigma_{ik}^2}(\|x_{i}-a_{k}\|-r_{ik})^2,
\end{equation}
and, for the bearings,~$f_u(x)$ is defined as
% \begin{equation}
%   \label{eq:noncvx-bearings}
%   \begin{split}
%     f_u(x) = &\sum_{i \sim j \in {\cal E}_u} \left(\kappa_{ij}u_{ij}^T\frac{x_{i}-x_{j}}{\|x_{i}-x_{j}\|}\right )\\
%     & + \sum_i \sum_{k \in {\cal U}_i} \left(\varkappa_{ik}v_{ik}^T\frac{x_{i}-a_{k}}{\|x_{i}-a_{k}\|}\right ).
% \end{split}
% \end{equation}
\begin{equation}
  \label{eq:noncvx-bearings}
  \sum_{i \sim j \in {\cal E}_u} \left(\kappa_{ij}u_{ij}^T\frac{x_{i}-x_{j}}{\|x_{i}-x_{j}\|}\right )
    + \sum_i \sum_{k \in {\cal U}_i} \left(\varkappa_{ik}v_{ik}^T\frac{x_{i}-a_{k}}{\|x_{i}-a_{k}\|}\right ).
\end{equation}
The unconstrained problem~\eqref{eq:noncvx-mle} is nonconvex due to
both terms~\eqref{eq:noncvx-ranges},~\eqref{eq:noncvx-bearings} and
difficult to solve globally. Function~$f$ in~\eqref{eq:noncvx-ranges}
is nonconvex because the argument of the square has a negative region
when~$\|x_{i}-x_{j}\| < d_{ij}$ (the same for the anchor
terms). Non-convexity of~$f_{u}$ stems from~$x_{i}-x_{j}$ appearing
nonlinearly in the denominator. We will overcome this difficulty by
relaxing the problem to a convex one, as presented in the next
section. Later, we will see in numerical results that the relaxation
retains good estimation accuracy.

\section{Convex relaxation}
\label{sec:cvx-relax}

Following~\cite{soares2014simple}, we rewrite each term
in~\eqref{eq:noncvx-ranges} as
\begin{equation}
  \label{eq:variational}
  (\|x_i-x_j\| - d_{ij})^2 = \min_{\|y_{ij}\| = d_{ij}} \|x_i-x_j-y_{ij}\|^2,
\end{equation}
where the constraint set represents a sphere centered at the origin
with radius~$d_{ij}$. 
% %
% \begin{figure}[!tb]
% \centering
% \includegraphics[width=0.6\columnwidth]{circle_variational_draft}%
% \caption{Variational representation of range-related terms.}
% \label{fig:circle_variational}
% \end{figure}
% %
When $y_{ij}$ is placed optimally on the circle with radius $d_{ij}$
its distance to $x_i$ is $\lvert \|x_i-x_j\| - d_{ij} \rvert$, as
intended. The auxiliary variable is readily worked out in closed form
as $y_{ij} = d_{ij}\frac{x_{i}-x_{j}}{\|x_{i}-x_{j}\|}$. Focusing on
inter-node terms only for clarity, we have for the hybrid ML problem
\begin{equation}
  \label{eq:ncvx-p}
  \begin{split}
    p_1 = & \min_{x,y} \sum_{i \sim j}\|x_i-x_j-y_{ij}\|^2 - \kappa_{ij}{u}_{ij}^T\frac{x_{i}-x_{j}}{\|x_{i}-x_{j}\|}\\
    &\subjto \: \|y_{ij}\| = d_{ij},
  \end{split}
\end{equation}
where~$y$ is the concatenation of~$\{y_{ij}, i \sim j\}$, constraints
are on all edges of~${\cal G}$, and~$p_1$ denotes the optimal value of the nonconvex
problem~\eqref{eq:noncvx-mle}. This can be equivalently written as
\begin{equation}
  \label{eq:ncvx-p1}
    \begin{split}
    p_1 = & \min_{x,y} \sum_{i \sim j}\|x_i-x_j-y_{ij}\|^2 - \frac{\kappa_{ij}}{d_{ij}}u_{ij}^Ty_{ij}\\
    &\subjto \: \|y_{ij}\| = d_{ij}, \: y_{ij} = d_{ij}\frac{x_{i}-x_{j}}{\|x_{i}-x_{j}\|},
  \end{split}
\end{equation}
% where~$p_1 = p$.
Note that the first constraint is redundant given the second one. We will now \emph{relax} our problem by dropping the
second constraint in~\eqref{eq:ncvx-p1}, obtaining
\begin{equation}
  \label{eq:ncvx-p2}
    \begin{split}
    p_2 = & \min_{x,y} \sum_{i \sim j}\|x_i-x_j-y_{ij}\|^2 - \tilde{u}_{ij}^Ty_{ij}\\
    &\subjto \: \|y_{ij}\| = d_{ij},
  \end{split}
\end{equation}
where~$\tilde{u}_{ij}=\frac{\kappa_{ij}}{d_{ij}}u_{ij}$. As the constraint set was enlarged, we have~$p_2 \leq p_1$.

%
% \begin{figure}[!tb]
% \centering
% \subfloat[Ranges only]{\includegraphics[width=0.5\columnwidth]{intersect_ranges_draft}%
% \label{fig:intersect_ranges}}
% \hfil
% \subfloat[Ranges and bearings]{\includegraphics[width=0.5\columnwidth]{intersect_hybrid_draft}%
% \label{fig:intersect_hybrid}}
% \caption{Relaxing the terms of the ML cost function.}
% \label{fig:intersect_ml}
% \end{figure}
%

\paragraph*{Disk relaxation}
\label{relax-disk}
Now we relax the constraint set from the sphere to the
ball~$\{y \colon \|y\| \leq d_{ij} \}$, its convex hull, to obtain an
approximation of the variational representation for range
terms~\eqref{eq:variational}
\begin{equation}
  \label{eq:variational_relaxed}
  \min_{\|y_{ij}\|  \leq d_{ij}} \|x_i-x_j-y_{ij}\|^2.
\end{equation}
%
% The variable~$y_{ij}$ at optimality is matching the
% difference vector~$x_i-x_j$, forcing this vector to have a norm as
% similar as possible to the acquired range measurement~$d_{ij}$.
As discussed in~\cite{soares2014simple}, replacing the
terms~\eqref{eq:variational} in the range-only cost
function~\eqref{eq:noncvx-ranges} with the modified
ones~\eqref{eq:variational_relaxed} is beneficial for outlier
rejection; if $y_{ij}$ can be placed anywhere on the ball, not just on
the border, this will limit the contribution of large disks created by
outliers with large values of $d_{ij}$. However, placing $x_i$ and
$y_{ij}$, $y_{il}$ anywhere inside the intersection area of such disks
will yield zero contribution to the cost.

Now consider the hybrid problem~\eqref{eq:ncvx-p2} after the same disk
relaxation of its constraint sets (relaxing~$\|y_{ij}\| = d_{ij}$
to~$\|y_{ij}\| \leq d_{ij}$). The newly added angular terms will break
the flatness of range contributions discussed previously, biasing the
$y$ variables back towards the borders of the disks along the
directions measured. Effectively, this formulation approximates the
intended behavior of the original one in~(\ref{eq:ncvx-p1}) with
equality constraints, while doing so in a soft way that 
preserves the ability to seamlessly reduce the impact of outliers in
range measurements.
The complete relaxed ML problem is as follows
\begin{equation}
  \label{eq:cvx}
  \begin{split}
     \minimize_{x,y,w} & \sum_{i \sim j}\|x_i-x_j-y_{ij}\|^2 - \tilde{u}_{ij}^Ty_{ij}\\
     & + \sum_{i \in {\cal V}, k \in {\cal A}_i}\|x_i-a_k-w_{ik}\|^2 - \tilde{v}_{ij}^Tw_{ik}\\
    \subjto & \: \|y_{ij}\| \leq d_{ij}, \: \|w_{ik}\| \leq r_{ik},
  \end{split}
\end{equation}
where~$w_{ik}$ have the corresponding role to~$y_{ij}$ regarding
anchor-node terms,
and~$\tilde{v}_{ik} = \frac{\varkappa_{ik}}{r_{ik}} v_{ik}$. This
non-standard relaxation is the main contribution of this work.

\section{Suboptimality analysis}
\label{sec:bound}

After presenting a convex relaxation to the nonconvex
problem~\eqref{eq:noncvx-mle} we now perform a tightness analysis. From
the derivation of our relaxation we know that if the optimal edge
variables~$y_{ij}^\star$, anchor-node variables~$w_{ik}^\star$, and
node positions~$x_i^\star$ obey the dropped equality constraints
$ y_{ij}^\star =
d_{ij}\frac{x_{i}^\star-x_{j}^\star}{\|x_{i}^\star-x_{j}^\star\|},
\qquad w_{ik}^\star =
r_{ik}\frac{x_{i}^\star-a_k}{\|x_{i}^\star-a_k\|}, $ then the
solution~$(x^\star, y^\star,w^\star)$ of problem~\eqref{eq:cvx} is
also the solution of the original nonconvex
problem~\eqref{eq:ncvx-p1}, considering anchors. We measure the
suboptimality in the optimization variables by the average
$p_1$-residual
\begin{multline}
  \label{eq:E1}
    E_{1} = \frac1{|{\cal E}|}\sum_{i \sim j} \left \|y_{ij}^\star - d_{ij}\frac{x_{i}^\star-x_{j}^\star}{\|x_{i}^\star-x_{j}^\star\|} \right \|\\
    + \sum_{i \in {\cal V}, k \in {\cal A}_i} \frac1{|{\cal A}_i|}
    \left\| w_{ik}^\star -
      r_{ik}\frac{x_{i}^\star-a_k}{\|x_{i}^\star-a_k\|} \right\|.
\end{multline}
A weaker, but rather useful result, is the verification of
$
  \|y_{ij}\| = d_{ij}$ and $ \|w_{ik}\| = r_{ik},
$
and the computation of the average $p_2$-residual
\begin{equation}
  \label{eq:E2}
  E_{2} = \frac1{|{\cal E}|}\sum_{i \sim j} |\|y^\star_{ij}\| - d_{ij}| +  \sum_{i \in {\cal V}, k \in {\cal A}_i} \frac1{|{\cal A}_i|} |\|w^\star_{ik}\| - r_{ik}|.
\end{equation}
Both results are important to understand how good our estimate for the
node positions~$x$ is.  We point out that, in the presence of noisy
measurements,~$E_1$ will not be zero, but in our simulations~$E_2$ is
indeed very close to zero. As the value of~$E_2$ is consistently very small in our numerical
experiments, the norms of edge variables~$y_{ij}$ and anchor-node
variables~$w_{ik}$ effectively equal the measured ranges. Thus,
it is also useful to consider the suboptimality angles defined by
% \begin{equation}
%   \label{eq:error-angles}
%   \begin{split}
%   \theta_{ij} & = \arccos\left\langle \frac{y_{ij}^\star}{\|y_{ij}^\star\|} , \frac{x_{i}^\star-x_{j}^\star}{\|x_{i}^\star-x_{j}^\star\|} \right \rangle, \\
%   \beta_{ik} & = \arccos\left\langle \frac{w_{ik}^\star}{\|w_{ik}^\star\|} , \frac{x_{i}^\star-a_{k}}{\|x_{i}^\star-a_{k}\|} \right \rangle,
% \end{split}
% \end{equation}
\begin{equation}
  \label{eq:error-angles}
  \theta_{ij}  = \arccos\left\langle \frac{y_{ij}^\star}{\|y_{ij}^\star\|} , \frac{x_{i}^\star-x_{j}^\star}{\|x_{i}^\star-x_{j}^\star\|} \right \rangle,
\end{equation}
where $\langle \cdot, \cdot \rangle$ denotes the usual inner product of two (unit-norm) vectors.  We also define~$\beta_{ik}$ similarly to~$\theta_{ij}$, but with node-anchor variables. Jointly with~$E_2$, these angles show how much our estimates deviate
from optimality.

\section{Numerical experiments}
\label{sec:experiments}

% Experiment setup?
To analyse performance and suboptimality, we randomly generated
geometric networks based on sensing ranges, and tested each network
for range localizability~\cite{AndersonShamesMaoFidan2010}, to
ascertain that there is no ambiguity in the solution space inherent to
network configurations. In the interest of visualization, we chose a
2D environment to perform our experiments. We stress, however, that
our algorithm is agnostic to the dimensionality of the ambient space.

\paragraph*{Problem size}
\label{prbm-sz}
We test our method on networks with $n=100$, and networks of $n=10$
nodes. The smaller sized networks are used for comparison with a
state-of-the-art SDP relaxation. Larger networks could not be solved
with a generic SDP solver.  Agents and anchors are randomly located in
a $7 \times 7\: m^2$ region, and, following the minimum number of
anchors allowed for range-only localization in 2D, we
set~$|{\cal A}| = 3$.  We emphasize that our method, minimizing a
quadratic over a convex set, practically can accommodate much larger
problem sizes than SDP-based formulations.

\paragraph*{Measurement data generation}
\label{data-model}
Range measurements are contaminated by noise with standard deviation
of 0.5~m, while bearing measurements are corrupted by noise with
standard deviation of~$2^\circ$. The concentration parameter
associated with each angular measurement is the inverse of the
variance in radians. These uncertainty values were drawn
from~\cite{PaullSaeediSetoLi2014}, regarding a relevant application of
hybrid localization algorithms: the underwater scenario.

\paragraph*{Simulation parameters}
\label{sim-pars}
The number of Monte Carlo (MC) trials, $M$, in each experiment, was
obtained by instantiating problems, running estimators, computing
metrics, and stopping whenever the running averages across MC
trials $ \langle H \rangle_M = \frac 1M \sum_{m=1}^M H_m $ were
sufficiently stable. Here,~$H$ stands for an error, for example,~$E_1$
in~\eqref{eq:E1}, $E_2$ in~\eqref{eq:E2} or the angles~$\theta_{ij}$
and~$\beta_{ik}$ in~\eqref{eq:error-angles}, computed from data of
MC trial~$m$.

\paragraph*{Tightness measures}
\label{sec:relax-bound}

We first check in simulation that the convex relaxation~\eqref{eq:cvx}
is tight regarding the nonconvex problem~\eqref{eq:ncvx-p2}. For this
experiment, the number of MC trials was~209.
\begin{figure}[tb]
  \centering
  \includegraphics[width=\columnwidth]{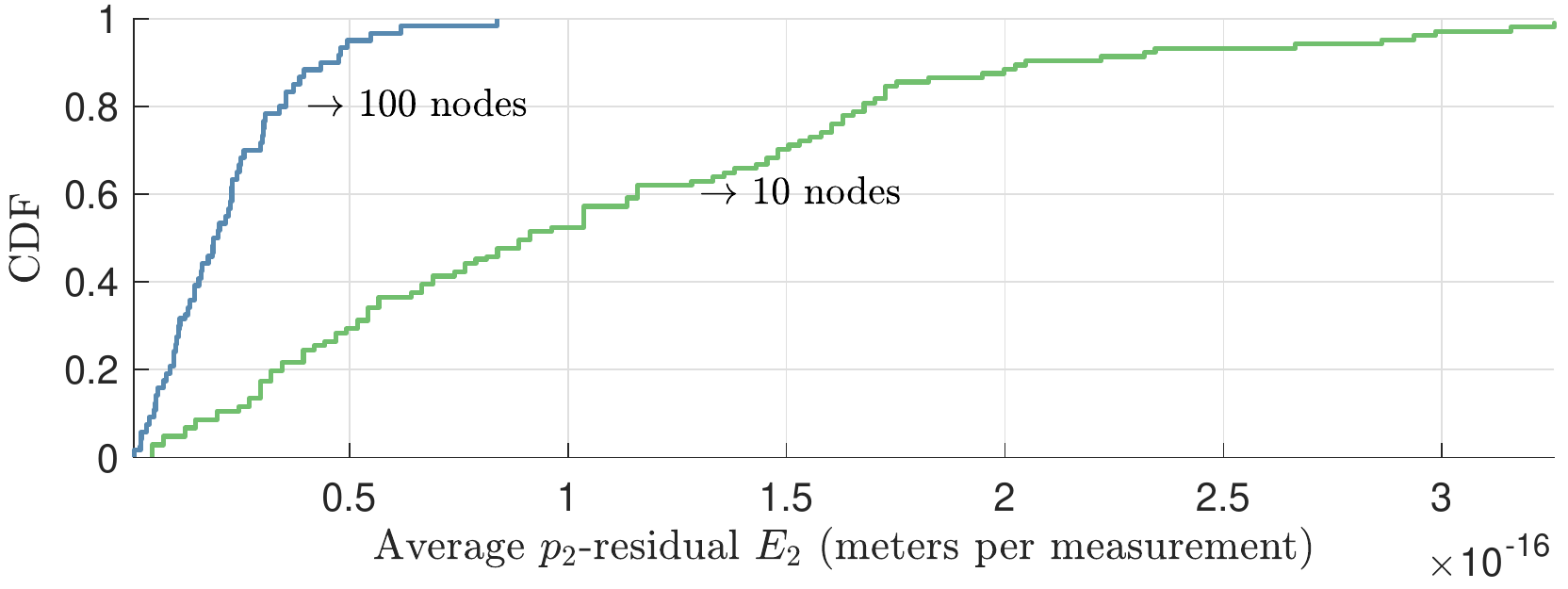}
  \caption[E2-100]{The empirical CDF shows that, for all MC
    trials, average~$p_2$-residuals~$E_2$ are below~$10^{-15}$. Thus,
    for all our experiments,~$\|y_{ij}^\star\| = d_{ij}$,
    and~$\|w_{ik}^\star\|= r_{ik}$ so, for all practical purposes,
    there is no suboptimality gap by approximating the
    nonconvex problem~\eqref{eq:ncvx-p2} with the convex
    relaxation~\eqref{eq:cvx}. We also observe that scaling the
    problem from 10 to 100 nodes slightly increases performance. In
    fact, due to having a denser network and more measurements ---
    albeit with more unknowns --- our estimator shows less violation
    of the nonconvex ML constraint dropped in~\eqref{eq:cvx}.}
  \label{fig:cdf-E2-100}
\end{figure}
The empirical Cumulative Distribution Function (CDF) of the ~$E_2$
residuals in Fig.~\ref{fig:cdf-E2-100} evidences that, for all MC
trials, the relaxation of the equality constraint on the edge and
node-anchor variables is tight, and that the solution
of~\eqref{eq:cvx} practically coincides with the solution
of~\eqref{eq:ncvx-p2}. This is a very interesting result reinforcing
the intuitive idea that if we add new independent measurements we
achieve better estimation.  Now we investigate $p_1$-residuals,
associated with dropping the constraints
$ y_{ij} = d_{ij}\frac{x_{i}-x_{j}}{\|x_{i}-x_{j}\|}$ and
$w_{ik} = r_{ik}\frac{x_{i}-a_{k}}{\|x_{i}-a_{k}\|}.  $
\begin{figure}[tb]
  \centering
  \includegraphics[width=\columnwidth]{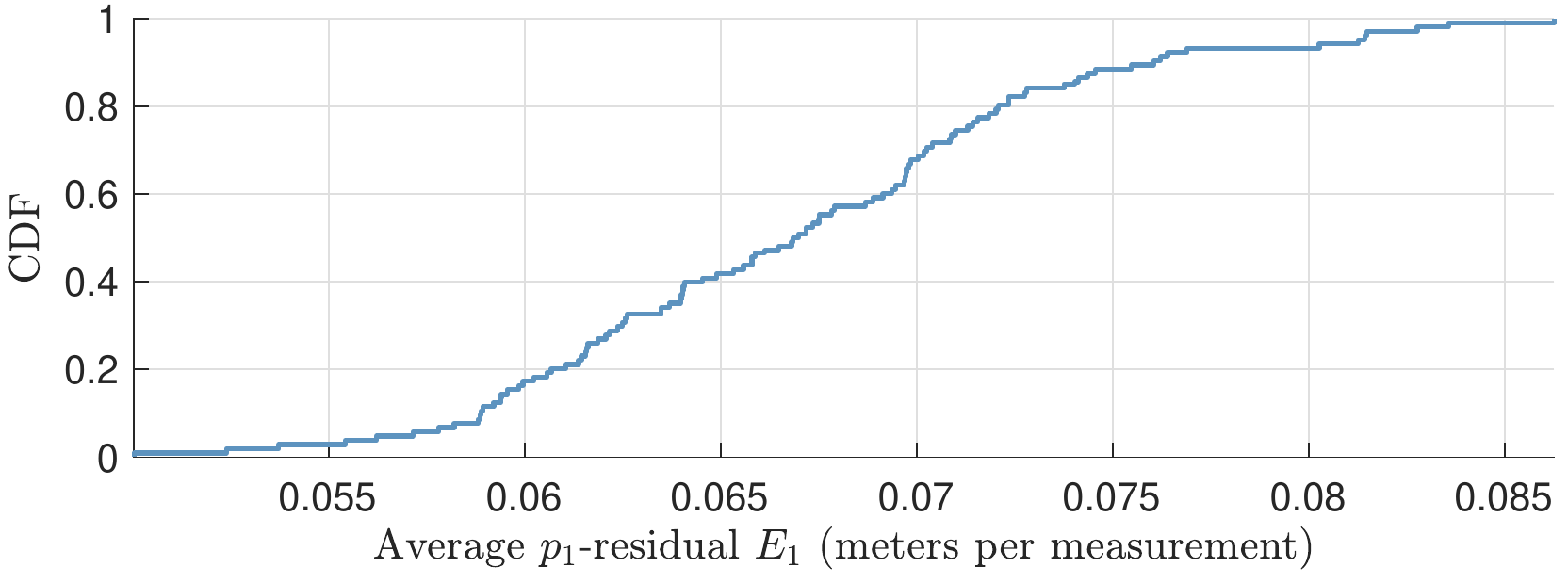}
  \caption[E1]{Empirical CDF for the average $p_1$-residual for
    networks with 10 nodes.  Approximating the nonconvex
    problem~\eqref{eq:ncvx-p1} with the nonconvex
    relaxation~\eqref{eq:ncvx-p2} represents an average error per
    measurement below 9 cm.}
  \label{fig:cdf-E1}
\end{figure}
Fig.~\ref{fig:cdf-E1} shows that the $p_1$-residual~$E_1$ is smaller
than range noise.
\begin{figure}[tb]
  \centering
  \includegraphics[width=\columnwidth]{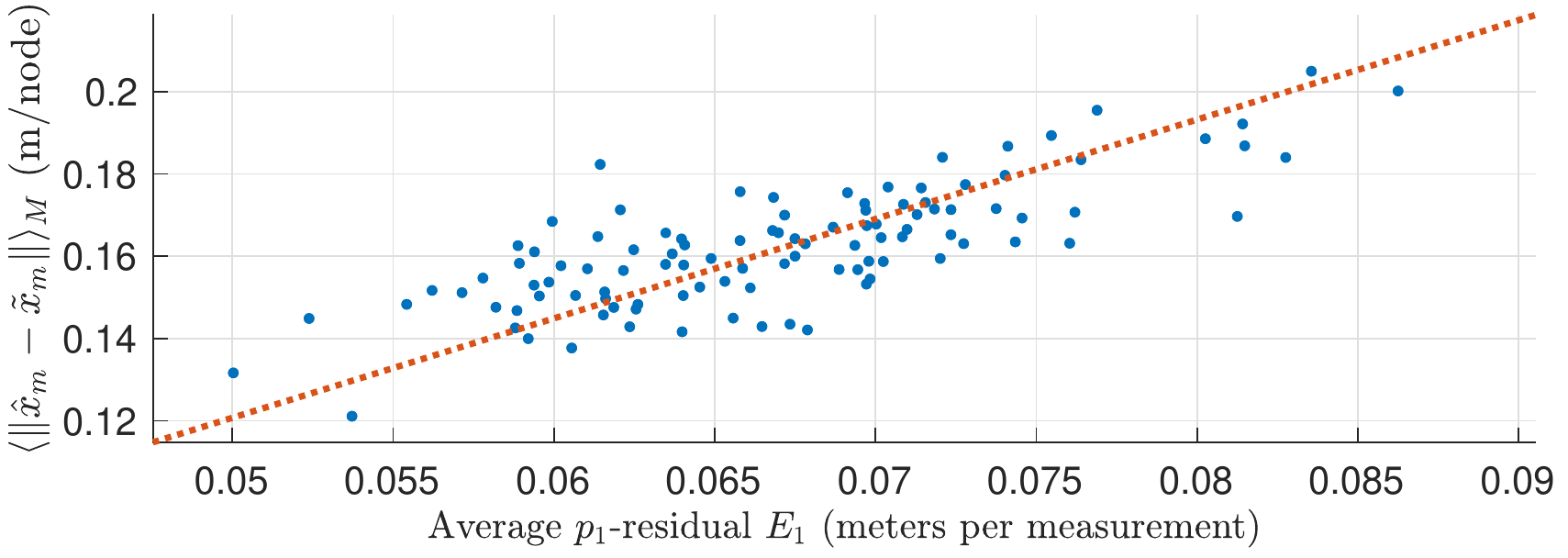}
  \caption[E1versusEstError]{There is a strong correlation between the
    $E_1$ residual error and position error. In MC trial~$m$,
    we compute the error of our estimate~$\hat{x}_m$ to the numerical
    minimizer of the nonconvex original
    problem~\eqref{eq:noncvx-mle},~$\tilde{x}_m$ when initialized on
    the ground truth pre-noise node positions.}
  \label{fig:EfminconVsE1}
\end{figure}
The approximation error~$E_1$ grows linearly with the error in the
estimates, as seen in Fig.~\ref{fig:EfminconVsE1}. In fact, when
plotted against the average error over MC trials between our
estimate~$\hat{x}_m$, and the numerical minimizer~$\tilde{x}_m$
of~\eqref{eq:noncvx-mle} initialized with the true positions before
noise addition, we see that the two errors are highly correlated. As
we have seen in Fig.~\ref{fig:cdf-E2-100}, there is negligible error
associated with the last relaxation. Whenever this is true, it is more
illuminating to observe the approximation error in terms of the
suboptimality angles~$\theta_{ij}$~\eqref{eq:error-angles}
and~$\beta_{ik}$.
\begin{figure}[tb]
  \centering
  \includegraphics[width=\columnwidth]{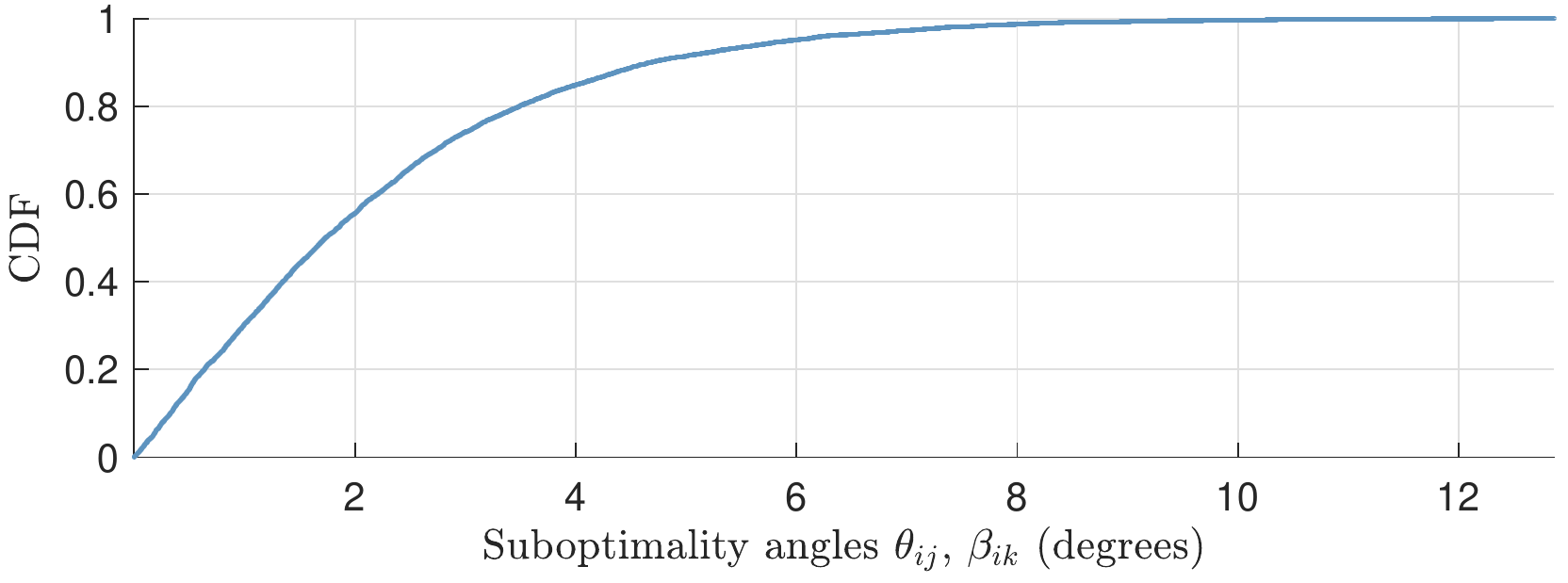}
  \caption[Angle error]{Empirical CDF for suboptimality
    angles~\eqref{eq:error-angles}. Angles are small. We
    observe that more than 80\% of the errors are below 4$^\circ$.  We
    can, then, conclude that the relaxation gives quite accurate
    results. The CDF shows more points than the previous figures
    because all angles are shown (they are not averaged).}
  \label{fig:cdf-angle-error}
\end{figure}
Fig.~\ref{fig:cdf-angle-error} evidences that the relaxation from
problem~\eqref{eq:ncvx-p1} to~\eqref{eq:ncvx-p2} does not incur a
large approximation error. We stress that the suboptimality angles
in~\eqref{eq:error-angles}, jointly with the verification that~$E_1$
is virtually zero, are intuitive metrics of how far is the optimum of
the true ML estimator for hybrid localization.

\paragraph*{Benchmark}
\label{sec:benchmark}
So far we have studied the performance of our relaxation with respect
to the nonconvex ML estimator, using our measures of suboptimality in
the solution. This section presents a comparison with a state of the
art method~\cite{Naseri2018}, using precisely the same data model as
our proposal. We used the generic solver \texttt{CVX}~\cite{cvx} to
obtain the SDP estimate.
\begin{figure}[tb]
  \centering
  \includegraphics[width=\columnwidth]{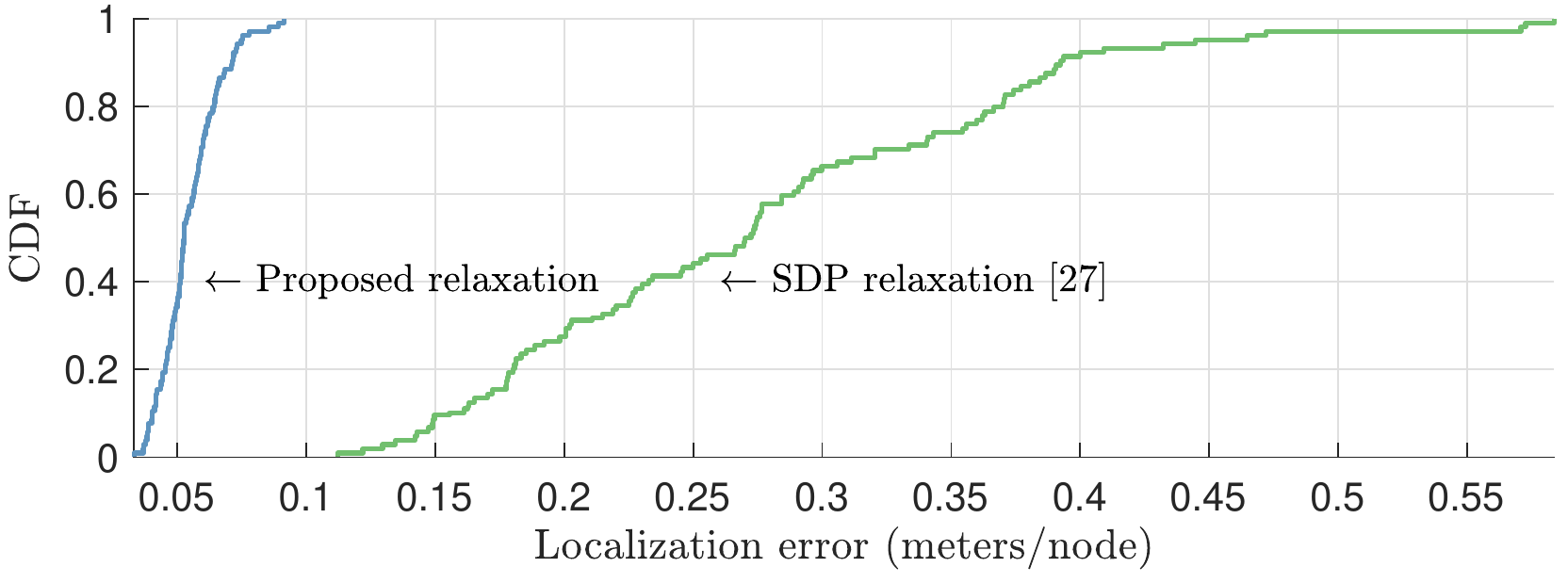}
  \caption[Error comparison with benchmark]{Empirical CDF of
    localization error for the proposed relaxation, compared with the SDP relaxation~\cite{Naseri2018}. Not only does our relaxation fare
    better in minimum, average, and maximum error per node, but the slope of
    the error CDF is also much steeper. This indicates low variance in
    the estimator performance. The median is one order of magnitude
    smaller for the proposed relaxation, and the maximum localization
    error is below~0.1~m, while the SDP method can surpass~0.55~m of
    localization error per node.}
  \label{fig:cdf-err-benchmark}
\end{figure}
This time we will measure performance based on the average
\emph{localization error} defined as
$e = \frac 1n \sum_{i} \| \hat{x}_i - x_i^\star\|$, where~$\hat{x}_i$
is the optimum of the position of node~$i$, given by one of the
relaxations, and~$x_i^\star$ is the true position, before measurement
noise.  Fig.~\ref{fig:cdf-err-benchmark} evidences the accuracy gain
in using the proposed approach. In our experiments, our approach
always yielded less than~0.1~m in localization error, while the SDP
relaxation resulted in a spreaded error from~0.1~m to~0.55~m. Summing
up, with one order of magnitude smaller error, our relaxation has less
variance and lower computational demands, offering even a practical
metric of suboptimality for our estimates.

\paragraph*{Scalability}
\label{sec:scalability}
Next, we ran 120 MC trials with similar random geometric networks, but
now with~100 nodes. We could not run the SDP relaxation for such large
networks, so our analysis will focus on the suboptimality measures
discussed in Section~\ref{sec:bound}.  The curve for 100 nodes in
Fig.~\ref{fig:cdf-E2-100} shows that, similarly to what we observed
previously, the relaxation of~(\ref{eq:ncvx-p2}) is tight.
\begin{figure}[tb]
  \centering
  \includegraphics[width=\columnwidth]{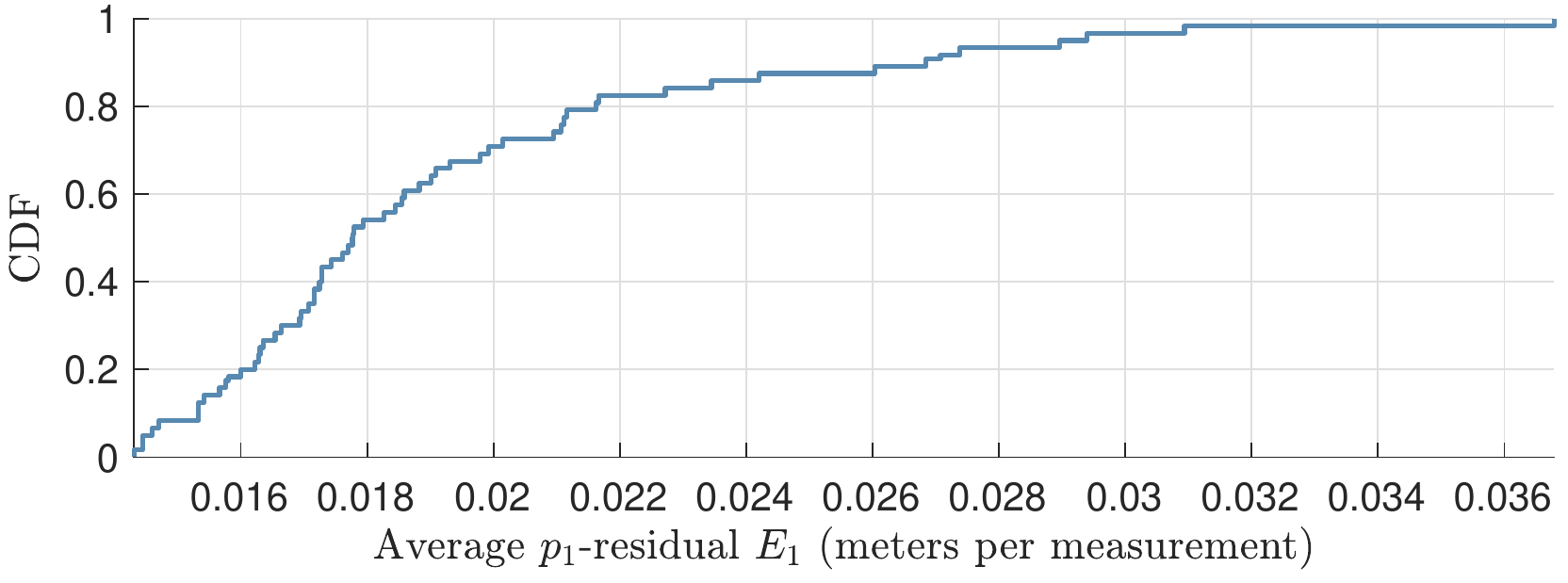}
  \caption[E1-100]{CDF of the average $p_1$ residual for the
    experiment with~100 nodes. The span of values has of~0.035~m per
    measurement, slightly better than the~0.085~m per measurement from
    the previous experiment in Fig.~\ref{fig:cdf-E1}.}
  \label{fig:cdf-E1-100}
\end{figure}
A similar observation can be drawn from Fig.~\ref{fig:cdf-E1-100}
regarding the more challenging of the two relaxations we performed.
We note that both of the residuals per measurement are smaller with
larger networks.

\paragraph*{\textbf{Conclusions}}
\label{sec:conclusions}
We presented a non-canonical relaxation of the hybrid
network localization problem, with excellent accuracy, and where
optimality certificates correlate with positioning error, informing on
the quality of the approximate solution.

% \appendices

% % you can choose not to have a title for an appendix
% % if you want by leaving the argument blank
% \section{}

% % use section* for acknowledgment
% \section*{Acknowledgment}


\ifCLASSOPTIONcaptionsoff
  \newpage
\fi



%\bibliographystyle{IEEEtran}
\bibliographystyle{IEEEtran}
\bibliography{IEEEabrv,biblos,refs_jp}

\begin{thebibliography}{10}
\providecommand{\url}[1]{#1}
\csname url@samestyle\endcsname
\providecommand{\newblock}{\relax}
\providecommand{\bibinfo}[2]{#2}
\providecommand{\BIBentrySTDinterwordspacing}{\spaceskip=0pt\relax}
\providecommand{\BIBentryALTinterwordstretchfactor}{4}
\providecommand{\BIBentryALTinterwordspacing}{\spaceskip=\fontdimen2\font plus
\BIBentryALTinterwordstretchfactor\fontdimen3\font minus
  \fontdimen4\font\relax}
\providecommand{\BIBforeignlanguage}[2]{{%
\expandafter\ifx\csname l@#1\endcsname\relax
\typeout{** WARNING: IEEEtran.bst: No hyphenation pattern has been}%
\typeout{** loaded for the language `#1'. Using the pattern for}%
\typeout{** the default language instead.}%
\else
\language=\csname l@#1\endcsname
\fi
#2}}
\providecommand{\BIBdecl}{\relax}
\BIBdecl

\bibitem{PaullSaeediSetoLi2014}
L.~Paull, S.~Saeedi, M.~Seto, and H.~Li, ``{AUV} navigation and localization: A
  review,'' \emph{IEEE Journal of Oceanic Engineering}, vol.~39, no.~1, pp.
  131--149, 2014.

\bibitem{Schneider2013}
D.~Schneider, ``You are here,'' \emph{{IEEE} Spectrum}, vol.~50, no.~12, pp.
  34--39, Dec. 2013.

\bibitem{Moustaka2019}
V.~Moustaka, A.~Vakali, and L.~G. Anthopoulos, ``A systematic review for smart
  city data analytics,'' \emph{ACM Computing Surveys (CSUR)}, vol.~51, no.~5,
  p. 103, 2019.

\bibitem{Witrisal2016}
K.~Witrisal, P.~Meissner, E.~Leitinger, Y.~Shen, C.~Gustafson, F.~Tufvesson,
  K.~Haneda, D.~Dardari, A.~F. Molisch, A.~Conti, and M.~Z. Win,
  ``High-accuracy localization for assisted living: {5G} systems will turn
  multipath channels from foe to friend,'' \emph{{IEEE} Signal Process. Mag.},
  vol.~33, no.~2, pp. 59--70, Mar. 2016.

\bibitem{Buehrer2018}
C.~E. O’Lone, H.~S. Dhillon, and R.~M. Buehrer, ``A statistical
  characterization of localization performance in wireless networks,''
  \emph{IEEE Transactions on Wireless Communications}, pp. 1--1, 2018.

\bibitem{Turjman2017}
F.~Al-Turjman, ``Positioning in the internet of things era: An overview,'' in
  \emph{2017 International Conference on Engineering and Technology (ICET)},
  Aug 2017, pp. 1--5.

\bibitem{Hossain2017}
M.~S. Hossain, ``Cloud-supported cyber–physical localization framework for
  patients monitoring,'' \emph{IEEE Systems Journal}, vol.~11, no.~1, pp.
  118--127, March 2017.

\bibitem{DeyEtAl2017}
N.~Dey, A.~S. Ashour, F.~Shi, and R.~S. Sherratt, ``Wireless capsule
  gastrointestinal endoscopy: Direction-of-arrival estimation based
  localization survey,'' \emph{IEEE Reviews in Biomedical Engineering},
  vol.~10, pp. 2--11, 2017.

\bibitem{AspnesErenGoldenbergAnderson2006}
J.~Aspnes, T.~Eren, D.~K. Goldenberg, A.~S. Morse, W.~Whiteley, Y.~R. Yang,
  B.~D. Anderson, and P.~N. Belhumeur, ``A theory of network localization,''
  \emph{IEEE Transactions on Mobile Computing}, vol.~5, no.~12, pp. 1663--1678,
  2006.

\bibitem{ji2004}
X.~Ji and H.~Zha, ``Sensor positioning in wireless ad-hoc sensor networks using
  multidimensional scaling,'' in \emph{IEEE INFOCOM 2004}, vol.~4.\hskip 1em
  plus 0.5em minus 0.4em\relax IEEE, 2004, pp. 2652--2661.

\bibitem{jiang2016}
W.~Jiang, C.~Xu, L.~Pei, and W.~Yu, ``Multidimensional scaling-based {TDOA}
  localization scheme using an auxiliary line.'' \emph{IEEE Signal Process.
  Lett.}, vol.~23, no.~4, pp. 546--550, 2016.

\bibitem{SoaresXavierGomes2014a}
C.~Soares, J.~Xavier, and J.~Gomes, ``Distributed, simple and stable network
  localization,'' in \emph{Signal and Information Processing (GlobalSIP), 2014
  IEEE Global Conference on}, Dec 2014, pp. 764--768.

\bibitem{Zhou2018}
S.~Zhou, N.~Xiu, and H.~Qi, ``A fast matrix majorization-projection method for
  penalized stress minimization with box constraints,'' \emph{IEEE Transactions
  on Signal Processing}, vol.~66, no.~16, pp. 4331--4346, Aug 2018.

\bibitem{SimonettoLeus2014}
A.~Simonetto and G.~Leus, ``Distributed maximum likelihood sensor network
  localization,'' \emph{Signal Processing, IEEE Transactions on}, vol.~62,
  no.~6, pp. 1424--1437, Mar. 2014.

\bibitem{soares2014simple}
C.~Soares, J.~Xavier, and J.~Gomes, ``Simple and fast convex relaxation method
  for cooperative localization in sensor networks using range measurements,''
  \emph{Signal Processing, IEEE Transactions on}, vol.~63, no.~17, pp.
  4532--4543, Sept 2015.

\bibitem{Piovesan2016}
N.~Piovesan and T.~Erseghe, ``Cooperative localization in wsns: a hybrid
  convex/non-convex solution,'' \emph{IEEE Transactions on Signal and
  Information Processing over Networks}, vol.~PP, no.~99, pp. 1--1, 2016.

\bibitem{ShiHeChenJiang2010}
Q.~Shi, C.~He, H.~Chen, and L.~Jiang, ``Distributed wireless sensor network
  localization via sequential greedy optimization algorithm,'' \emph{Signal
  Processing, IEEE Transactions on}, vol.~58, no.~6, pp. 3328 --3340, June
  2010.

\bibitem{Saab2016}
K.~K. Saab and S.~S. Saab, ``Application of an optimal stochastic
  {Newton-Raphson} technique to triangulation-based localization systems,'' in
  \emph{2016 IEEE/ION Position, Location and Navigation Symposium (PLANS)},
  April 2016, pp. 981--986.

\bibitem{Agiwal2016}
M.~Agiwal, A.~Roy, and N.~Saxena, ``Next generation {5G} wireless networks: A
  comprehensive survey,'' \emph{IEEE Communications Surveys Tutorials},
  vol.~18, no.~3, pp. 1617--1655, thirdquarter 2016.

\bibitem{Shen2010}
Y.~Shen and M.~Z. Win, ``Fundamental limits of wideband localization — {P}art
  {I}: {A} general framework,'' \emph{{IEEE} Trans. Inf. Theory}, vol.~56,
  no.~10, pp. 4956--4980, Oct. 2010.

\bibitem{Patwari2005}
N.~Patwari, J.~N. Ash, S.~Kyperountas, A.~O. Hero, R.~L. Moses, and N.~S.
  Correal, ``Locating the nodes: cooperative localization in wireless sensor
  networks,'' \emph{IEEE Signal processing magazine}, vol.~22, no.~4, pp.
  54--69, 2005.

\bibitem{Biswas2005}
P.~Biswas, H.~Aghajan, and Y.~Ye, ``Integration of angle of arrival information
  for multimodal sensor network localization using semidefinite programming,''
  in \emph{Proceedings of the 39th Annual Asilomar Conference on Signals,
  Systems, and Computers}, Pacific Grove, CA, USA, 2005.

\bibitem{eren2011}
T.~Eren, ``Cooperative localization in wireless ad hoc and sensor networks
  using hybrid distance and bearing (angle of arrival) measurements,''
  \emph{{EURASIP} Journal on Wireless Communications and Networking}, vol.
  2011, no.~1, p.~72, 2011.

\bibitem{FerreiraGomesCosteira2015}
B.~Q. Ferreira, J.~Gomes, and J.~P. Costeira, ``A unified approach for hybrid
  source localization based on ranges and video,'' in \emph{2015 IEEE
  International Conference on Acoustics, Speech and Signal Processing
  (ICASSP)}, April 2015, pp. 2879--2883.

\bibitem{FerreiraGomesSoaresCosteira2016}
B.~Ferreira, J.~Gomes, C.~Soares, and J.~P. Costeira, ``Collaborative
  localization of vehicle formations based on ranges and bearings,'' in
  \emph{2016 IEEE Third Underwater Communications and Networking Conference
  (UComms)}, Aug 2016, pp. 1--5.

\bibitem{Naseri2018}
H.~Naseri and V.~Koivunen, ``Convex relaxation for maximum-likelihood network
  localization using distance and direction data,'' in \emph{2018 IEEE 19th
  International Workshop on Signal Processing Advances in Wireless
  Communications (SPAWC)}, June 2018, pp. 1--5.

\bibitem{Ferreira2018}
B.~Q. Ferreira, J.~Gomes, C.~Soares, and J.~P. Costeira, ``{FLORIS} and
  {CLORIS}: Hybrid source and network localization based on ranges and video,''
  \emph{Signal Processing}, vol. 153, pp. 355--367, 2018.

\bibitem{wu2019cooperative}
Y.~Wu, B.~Peng, H.~Wymeersch, G.~Seco-Granados, A.~Kakkavas, M.~H.~C. Garcia,
  and R.~A. Stirling-Gallacher, ``Cooperative localization with angular
  measurements and posterior linearization,'' \emph{arXiv preprint
  arXiv:1907.04700}, 2019.

\bibitem{kay1993}
S.~M. Kay, \emph{Fundamentals of Statistical Signal Processing - Estimation
  Theory}.\hskip 1em plus 0.5em minus 0.4em\relax Prentice Hall, 1993, vol.~I.

\bibitem{OguzGomesXavierOliveira2011}
P.~O\u{g}uz-Ekim, J.~Gomes, J.~Xavier, and P.~Oliveira, ``Robust localization
  of nodes and time-recursive tracking in sensor networks using noisy range
  measurements,'' \emph{Signal Processing, IEEE Transactions on}, vol.~59,
  no.~8, pp. 3930 --3942, Aug. 2011.

\bibitem{Tomic2019}
S.~Tomic, M.~Beko, and M.~Tuba, ``A linear estimator for network localization
  using integrated {RSS} and {AoA} measurements,'' \emph{IEEE Signal Processing
  Letters}, vol.~26, no.~3, pp. 405--409, 2019.

\bibitem{Tomic2017}
S.~Tomic, M.~Beko, R.~Dinis, and J.~Gomes, ``Target tracking with sensor
  navigation using coupled {RSS} and {AOA} measurements,'' \emph{Sensors},
  vol.~17, no.~11, p. 2690, 2017.

\bibitem{AndersonShamesMaoFidan2010}
B.~D.~O. Anderson, I.~Shames, G.~Mao, and B.~Fidan, ``Formal theory of noisy
  sensor network localization,'' \emph{{SIAM} Journal on Discrete Mathematics},
  vol.~24, no.~2, pp. 684--698, 2010.

\bibitem{cvx}
M.~Grant and S.~Boyd, ``{CVX}: Matlab software for disciplined convex
  programming, version 1.21,'' \url{http://cvxr.com/cvx}, Apr. 2011.

\end{thebibliography}

\end{document}